\documentclass[aps,prb,twocolumn,superscriptaddress,showpacs]{revtex4}
\usepackage{graphics}
\usepackage{graphicx}
\usepackage[english]{babel}
\usepackage{color}

\begin{document}

\title{Geometrical confinement effects in layered mesoscopic vortex-matter}

\author{N. R. Cejas Bolecek}
\affiliation{Low Temperature Division, Centro At\'{o}mico
Bariloche, CNEA, Argentina}

\author{M. I. Dolz}
\affiliation{Universidad Nacional de San Luis, San Luis,
Argentina}

\author{A. Kolton}
\affiliation{Solid State Theory group, Centro At\'{o}mico
Bariloche, CNEA, Argentina}

\author{H. Pastoriza}
\affiliation{Low Temperature Division, Centro At\'{o}mico
Bariloche, CNEA, Argentina}

\author{C. J. van der Beek}
\affiliation{Laboratoire des Solides Irradi\'{e}s, Ecole
Polytechnique, Palaiseau, France}

\author{M. Konczykowski}
\affiliation{Laboratoire des Solides Irradi\'{e}s, Ecole
Polytechnique, Palaiseau, France}

\author{M. Menghini}
\affiliation{Katholieke Universitat Leuven, Leuven, Belgium}

\author{G. Nieva}
\affiliation{Low Temperature Division, Centro At\'{o}mico
Bariloche, CNEA, Argentina}

\author{Y. Fasano}
\affiliation{Low Temperature Division, Centro At\'{o}mico
Bariloche, CNEA, Argentina}

\date{\today}



\begin{abstract}
We study geometrical confinement effects in
Bi$_{2}$Sr$_{2}$CaCu$_{2}$O$_{8 +\delta}$ mesoscopic vortex-matter
with edge-to-surface ratio of $7-12$\,\%. Samples have in-plane
square and circular edges, 30\,$\mu$m widths, and $\sim 2\,\mu$m
thickness. Direct vortex imaging reveals the compact planes of the
structure align with the sample edge by introducing topological
defects. The defects density is larger for circular than for
square edges. Molecular dynamics simulations suggest this density
is not an out-of-equilibrium property but rather determined by the
geometrical confinement.

\end{abstract}

\maketitle

\section{Introduction}

Understanding the confinement effects introduced by sample
geometry is crucial for characterizing the static and dynamic
properties of mesoscopic vortex matter. This subject was actively
investigated for low-temperature superconductors with dimensions
comparable or smaller than  coherence length or penetration depth,
$\lambda$
\cite{Moshchalkov95,Geim97,Schweigert98b,Palacios98,Bruyndoncx99,Cabral04}.
Mesoscopic vortex matter in these materials have structural
properties strongly influenced by the geometry of the specimens
\cite{Schweigert98b}, in contrast with results in macroscopic
samples for several compounds
\cite{Fasano1999,Menghini2003,Petrovic2009,Demirdis}. Confinement
effects are in competition with inter-vortex interaction that
increases with field and temperature. Materials with an important
electronic anisotropy such as layered high-$T_{\rm c}$'s have
quite a large value of $\lambda$ and then inter-vortex
interactions become more relevant.

Due to the technical difficulties for fabricating micron-sized
samples of layered high-$T_{\rm c}$'s complex oxides,
 there are few works in the literature investigating the effect of confinement in
 vortex matter nucleated in these materials
\cite{Wangprb6502}. In this work we study this issue in the
paradigmatic Bi$_{2}$Sr$_{2}$CaCu$_{2}$O$_{8 + \delta}$ compound
that presents a rich vortex phase diagram governed by thermal
fluctuations and extremely anisotropic magnetic properties. In
this compound, the phase diagram of macroscopic as well as
mesoscopic \cite{Koncykowski2012} vortex matter is dominated by a
first-order transition \cite{Pastoriza94,Zeldov95} between a solid
phase at low temperatures and a liquid \cite{Nelson1988} or
decoupled gas \cite{Glazman1991,Pastoriza1995} of pancake vortices
at high temperatures. The  vortex solid phase of macroscopic
 samples presents quasi
long-range positional order \cite{Fasano2005}.

Here we report on the structural properties of the mesoscopic
vortex solid nucleated in Bi$_{2}$Sr$_{2}$CaCu$_{2}$O$_{8 +
\delta}$ at low fields and with single-vortex resolution. We study
both, experimentally and with simulations, the effect of
confinement and inter-vortex interactions for samples with square
and circular edges experimental.

\section{Methods}

We engineered micron-sized superconducting samples from bulk
Bi$_2$Sr$_2$CaCu$_2$O$_{8+\delta}$ crystals ($T_{\rm c} = 89\,$K).
We fabricated circular and square samples with typical dimensions
of $30\,\mu$m   by means of optical lithography and subsequent
physical ion-milling of the negative of the samples
\cite{Moira2010}. Freestanding $2\,\mu$m thick disks and cuboids
are obtained after cleaving the towers resulting from milling.

\begin{figure}[bbb]
\begin{center}
\includegraphics[width=0.5\textwidth]{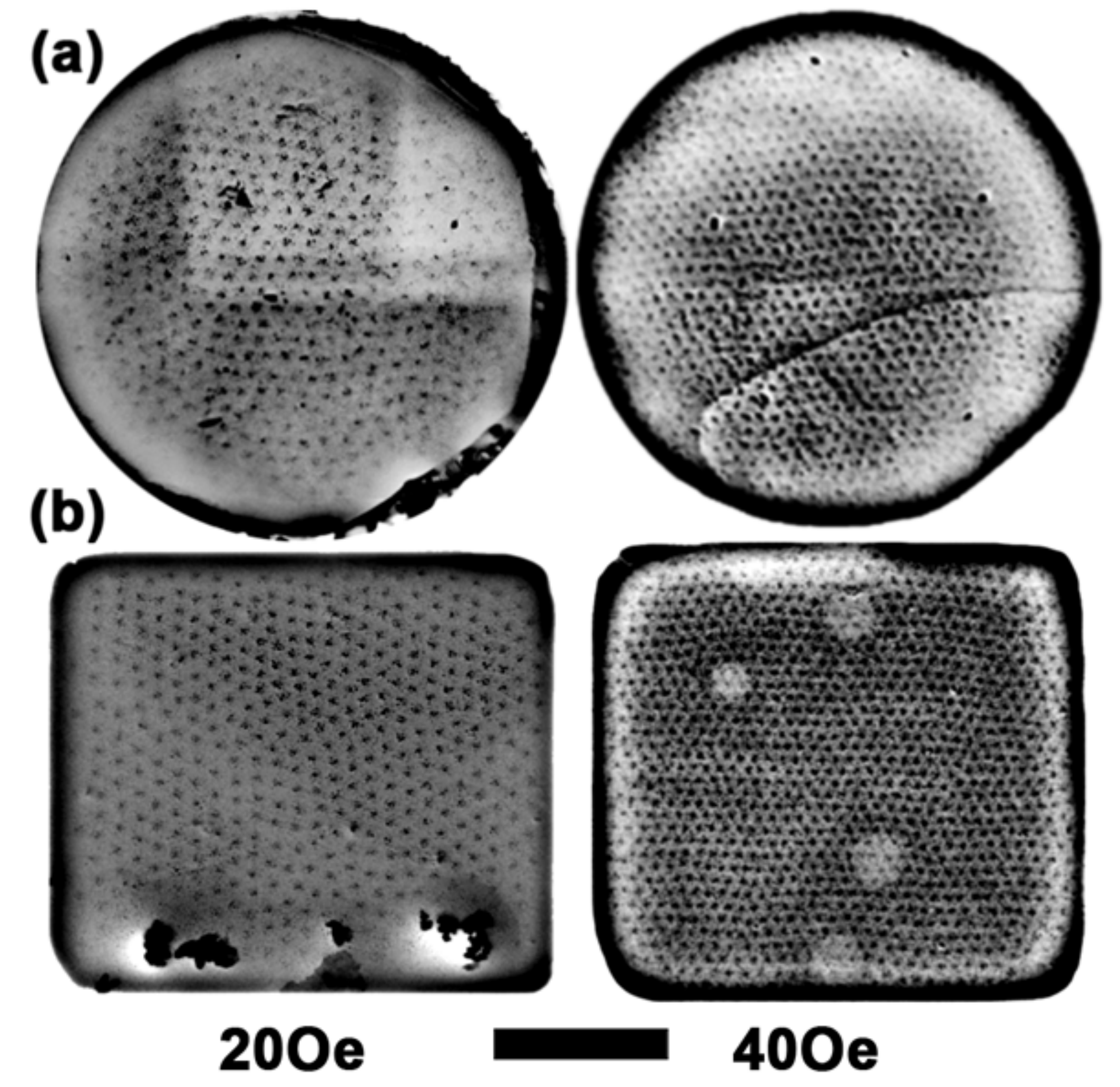}
\end{center}
\caption{\label{figure1} Magnetic decoration images of the
mesoscopic vortex matter nucleated in field-cooling processes in
micron-sized Bi$_2$Sr$_2$CaCu$_2$O$_{8+\delta}$ samples. (a) Disk
with 30\,$\mu$m diameter and (b) cuboid with 30\,$\mu$m sides
length, all samples with 2\,$\mu$m thickness. Magnetic decorations
were performed at 4.2\,K and at applied fields of 20 (left panel)
and 40\,Oe (right panel). The  scale-bar corresponds to
$10\,\mu$m. \label{fig:magdec30um}}
\end{figure}

We directly imaged  the solid vortex phase with single-vortex
resolution by means of magnetic decoration experiments
 performed at $4.2\,$K after field-cooling \cite{Fasano2003}. In
these experiments the evaporated magnetic nanoparticles land in
the sample surface at the places where the gradient of local
inductance is maximum, therefore decorating the vortex positions.
The imaged structure corresponds to the vortex solid frozen at the
temperature at which pinning sets in, $T_{\rm freez} \sim T_{\rm
irr}$ \cite{Fasano1999}, of the order of $90-87\,$K for the
low-fields studied here \cite{Dolz2014}. Decreasing the sample
size down to microns does not significantly affect the value of
$T_{\rm irr}$ \cite{Koncykowski2012}.

We also performed molecular dynamics simulations of
two-dimensional vortex matter  in order to emulate the
experimentally-observed  vortex matter structural properties
\cite{Kolton2000}. We studied the case of 30\,$\mu$m diameter
disks and focused on the density of topological defects when
varying the simulation cooling-rate.

\section{Results and discussion}

\begin{figure}[bbb]
\begin{center}
\includegraphics[ width=0.5\textwidth]{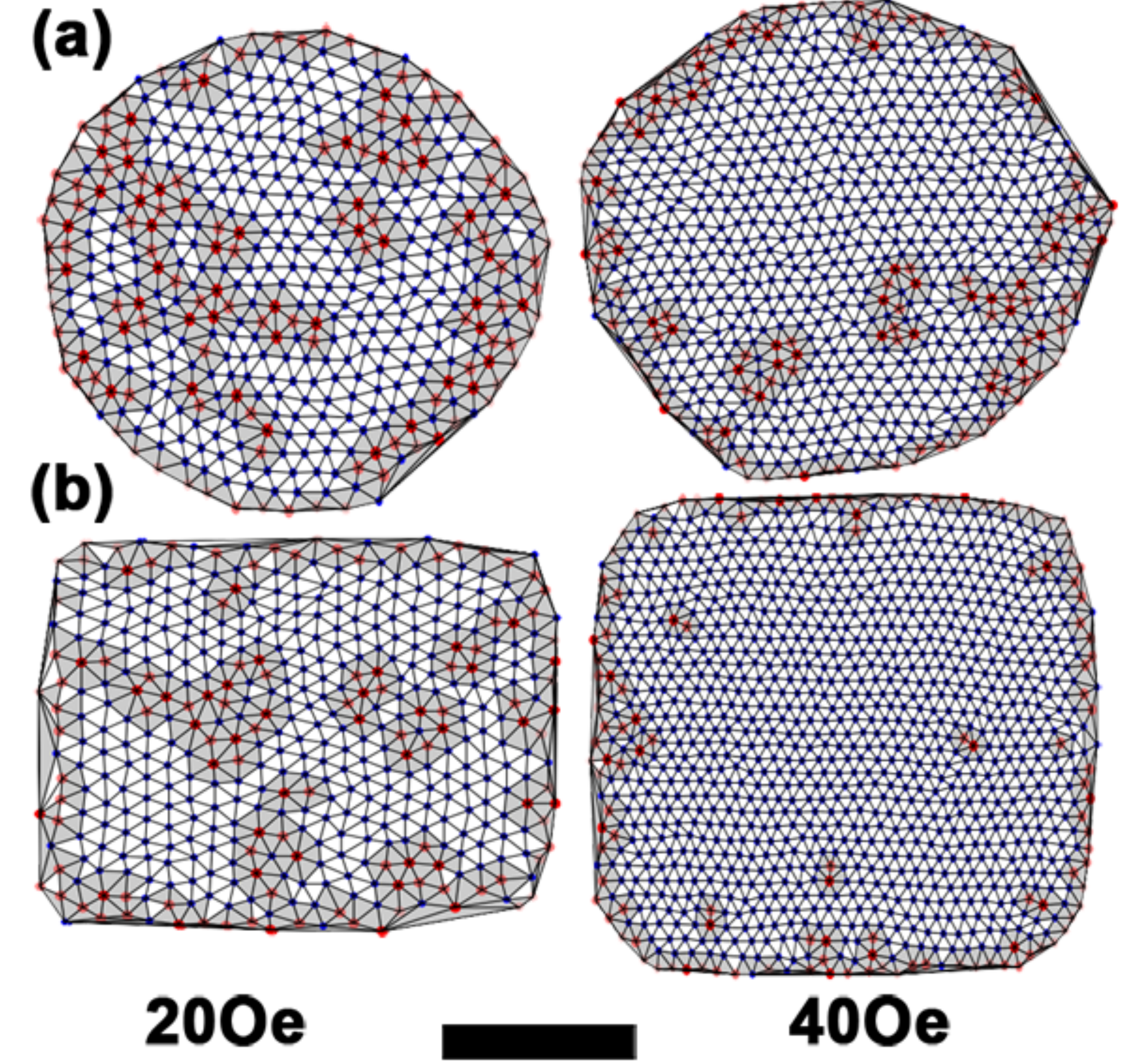}
\end{center}
\caption{\label{figure2} Delaunay triangulations of the mesoscopic
vortex matter shown in Fig.\,\ref{figure1}. Disclinations are
highlighted in gray; sixfold(non-sixfold)-coordinated vortices are
indicated in blue(red).  The  scale-bar corresponds to
$10\,\mu$m.}
\end{figure}

Figure\,\ref{figure1} shows snapshots of the mesoscopic vortex
structure nucleated in the disk and cuboid
Bi$_2$Sr$_2$CaCu$_2$O$_{8+\delta}$ samples after field-cooling
down to 4.2\,K at applied fields of 20 and 40\,Oe. The local
induction calculated as the number of vortices times the flux
quantum yields $B \sim 0.75 H$, a reduction due to demagnetizing
effects (aspect ratio of the disks and cuboids of $\sim 15$). The
direction of the compact planes of the vortex structure in
micron-sized specimens is affected by the confinement effect
introduced by the edges of the samples, particularly in the case
of the outer shells of vortices. This finding is in contrast to
observations in macroscopic samples \cite{Fasano1999}. The
alignment  is more evident in the Delaunay triangulations of
Fig.\,\ref{figure2}, a well known geometrical algorithm that
determines the first-neighbors for every  vortex in the structure
\cite{Fasano2005}. First-neighbors vortices are bounded with lines
and non-sixfold coordinated ones are highlighted in grey. In the
case of the cuboid samples, irrespective of the vortex density
(447 \textit{vs.} 1092), one of the compact planes of the
structure is parallel to the sample edge. For the disks, only a
few outer shells of vortices mimic the sample edges, the number
depending inversely with the vortex density. Towards the center of
the sample, a rather ordered vortex crystallite is formed with the
compact planes having no register with the sample edges. The
transition between the orientation of the outer and inner shells
is done via  the plastic deformations entailed by topological
defects.

\begin{figure}[bbb]
\begin{center}
\includegraphics[ width=0.5\textwidth]{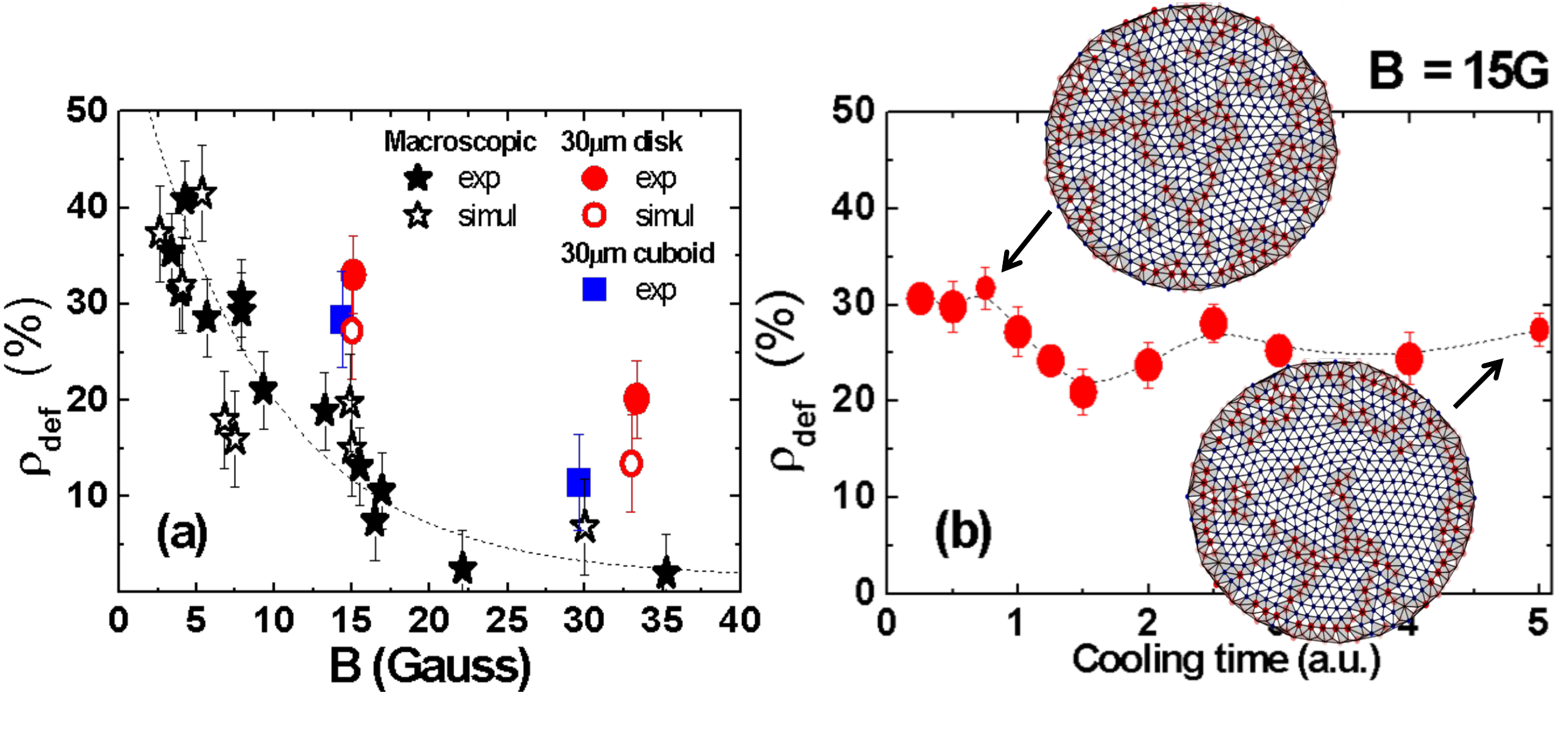}
\end{center}
\caption{\label{figure3} Density of topological defects
(non-sixfold coordinated vortices) in mesoscopic and macroscopic
Bi$_2$Sr$_2$CaCu$_2$O$_{8+\delta}$ vortex matter. (a) Experimental
data as a function of applied field for the mesoscopic vortex
structures nucleated in the 30\,$\mu$m disk and cuboid, and
macroscopic vortex matter. (b) Results from molecular dynamics
simulations as a function of the simulations relaxation time in
the case of a 30\,$\mu$m disk and $B=15$\,Gauss.}
\end{figure}

For the vortex structure studied here these topological defects
are generally disclinations, namely vortices with five or seven
first-neighbors, and pairs of them or screw-dislocations
associated to an extra plane of vortices. For example, isolated
dislocations are observed in the middle of the vortex structure
nucleated in the cuboid at an applied field of 40\,Oe. The density
of  non-sixfold coordinated vortices,  $\rho_{\rm def}$, strongly
depends on the local induction. In the case of macroscopic
Bi$_2$Sr$_2$CaCu$_2$O$_{8+\delta}$ vortex matter, $\rho_{\rm def}$
decreases exponentially up to 20\,Gauss and then saturates around
2\% as shown in Fig.\,\ref{figure3} (a). This is due to the
enhancement of inter-vortex interaction on increasing field. This
magnitude follows the same $B$-evolution for mesoscopic
 vortex matter but is at least
50\% larger  than for bulk samples. In addition, $\rho_{\rm def}$
is always larger in  disks than in  cuboids for roughly the same
vortex density. This can be explained by considering that aligning
a compact plane of vortices with the edges of a cuboid does not
imply to change the orientational order of the structure whereas
in order to do so in a disk the vortex planes have to bend.

In the case of macroscopic samples, it has been proved that the
structure observed by means of field-cooling decorations at
4.2\,K, and therefore its $\rho_{\rm def}$, is quite close to the
equilibrium \cite{Fasano2005}. The possibility of the increase on
the $\rho_{\rm def}$ on decreasing the system size being an
out-of-equilibrium phenomena can not be discarded. Therefore we
performed molecular dynamics simulations of the mesoscopic vortex
matter nucleated in a 30\,$\mu$m disk with a density of 15\,Gauss
in order to test this possibility. In particular, we performed
tests on the dependence of $\rho_{\rm def}$ with the cooling rate,
inversely proportional to the time allowed to the system to relax.
First, we performed simulations in a macroscopic sample in order
to find the pinning magnitude that has to be considered in order
to reproduce the observed $\rho_{\rm def}$. Then we used this
magnitude of pinning to perform simulations in micron-sized
samples. The results of $\rho_{\rm def}$ as a function  of the
cooling rate, see Fig.\,\ref{figure3} (b), indicate that the
observed experimental values correspond to the case of large
relaxation times. Therefore, we can ascertain that the amount of
topological defects observed in the experiments in micron-sized
samples is not an out-of-equilibrium feature and that the observed
structure is quite close to the equilibrium.

Having this certainty in mind, we tried to estimate the
geometrical confinement energy induced by the sample edges. In the
case of bulk samples, the mean value of the inter-vortex
interaction energy distribution is slightly shifted upward with
respect to the value for a perfect Abrikosov lattice with the same
vortex density \cite{Demirdis}. Since after a field-cooling
process vortices are close to equilibrium, this shifting can only
be accounted by the effect of bulk pinning \cite{Demirdis}. This
can be expressed, by unit length, as $<\epsilon_{\rm int}>^{\rm b}
- \epsilon_{\rm Abr} = \epsilon_{\rm p}^{\rm b}$, where $\rm b$
stands for the bulk sample and $<\epsilon_{\rm int}>$ is the mean
value of a distribution of inter-vortex interaction energies,
$\epsilon_{\rm Abr}$ the value of the inter-vortex interaction
energy in a perfect Abrikosov lattice (a delta-function), and
$\epsilon_{\rm p}$ the pinning energy. In the case of mesoscopic
vortex matter, an extra term enters into the energy-balance,
namely the confinement energy $\varepsilon_{\rm conf}$ and
therefore $<\epsilon_{\rm int}>^{\rm meso} - \epsilon_{\rm Abr} =
\epsilon_{\rm p}^{\rm meso} + \varepsilon_{\rm conf}$, where $\rm
meso$ means for the case of mesoscopic vortex matter and the
energies are noted similarly as in the previous case. Therefore,
one can have access to an estimation of the confinement energy in
mesoscopic vortex matter just by assuming that the pinning
magnitude is the same irrespective of the sample size, and then
$\varepsilon_{\rm conf} = <\epsilon_{\rm int}>^{\rm meso} -
<\epsilon_{\rm int}>^{\rm b}$.

The inter-vortex interaction energy per unit length depends on the
inter-vortex distances $r_{ij}$, and for a vortex $i$ has a value
$\epsilon_{\rm int}^{i} = \sum_{j} 2 \epsilon_{0}
K_{0}(r_{ij}/\lambda)$, with the sum over neighbor-vortices $j$,
$\epsilon_{0}\propto \lambda^{2}$ the vortex line tension, and
$K_{0}$ the zeroth-order modified Bessel function. In real cases
this magnitude is spatially inhomogeneous due to the elastic and
plastic deformations of the structure, and therefore there is a
distribution of $\epsilon_{\rm int}^{i}$ with an almost-Gaussian
shape \cite{Demirdis}. Only in the case of an ideal Abrikosov
lattice this magnitude is space invariant and its distribution is
a delta function. We have performed inter-vortex
energy-distribution calculations in vortex structures observed by
magnetic decoration in the macroscopic samples from which were
engineered the disks and cuboids. We also performed the same
calculations in the mesoscopic vortex matter nucleated in the
disks and cuboids at both applied fields. Irrespective of the
field, the mean values of $<\epsilon_{\rm int}>^{\rm meso}$ are
always larger than in the case of macroscopic vortex matter, what
can be reasonably ascribed to the extra deformations introduced by
the larger amount of topological defects nucleated in the
micron-sized samples. In accordance with this reasoning, the
$<\epsilon_{\rm int}>^{\rm meso}$ value is always larger for the
structure nucleated in the disks than in the cuboids by a 6-9\% on
increasing field. Therefore the estimated confinement energy is in
the case of disks equal to $1\pm 0.1 \times 10^{-8}$\,erg/cm$\sim
0.13 \epsilon_{0}$, larger than in the case of cuboids, $0.8\pm
0.1 \times 10^{-8}$\,erg/cm$\sim 0.11 \epsilon_{0}$.

\section{Conclusions}

The edges of the samples do produce a geometrical confinement
effect in mesoscopic vortex matter that is put in evidence by the
orientation of the outer shells of vortices with compact planes
parallel to the edges what produces a concomitant increase of the
density of topological defects. By means of molecular dynamics
simulations we show that the density of defects found
experimentally is not an out-of-equilibrium feature but rather the
effect introduced by geometrical confinement. By means of
differences in the mean value of the inter-vortex interaction of
the mesoscopic and macroscopic vortex structures we are able to
quantify the confinement energy per unit length. We find that is
just $0.11-0.13$\,$\epsilon_{0}$, the larger value in the case of
disks than cuboid geometries.

\end{document}